\documentclass{article}
\usepackage{amssymb}
\usepackage[utf8]{inputenc}
\usepackage[normalem]{ulem}
\usepackage{authblk}
\usepackage{setspace}
\usepackage[margin=1.25in]{geometry}
\usepackage{graphicx}
\graphicspath{ {./figures/} }
\usepackage{subcaption}
\usepackage{amsmath}
\usepackage{lineno}
\usepackage{comment}
\usepackage{float}
\usepackage{textcomp}
\usepackage[usenames,dvipsnames]{color}
\usepackage{hyperref}
\DeclareUnicodeCharacter{2212}{-}
\DeclareUnicodeCharacter{0308}{\"}
\usepackage[numbers,sort&compress]{natbib}
\usepackage{titlesec}
\titlelabel{\thetitle.\quad}

\title{\bf Spin–Orbit Coupling Effects on the Structural and Electronic Properties of Planar Pentagonal p-MS$_{2}$ (M = Si, Ge, and Pb)}
\author[1,$\dagger$]{Phuc-Dang Truong}
\author[1]{Cao-Huu-Tai Nguyen}
\author[1]{Nguyen-Bao-Tran Ngo}
\author[1]{Khanh-Van Huynh}
\author[3]{J\'{a}n Min\'{a}r}
\author[2,*]{Worawat Meevasana}
\author[1,*,$\dagger$]{Yen-Mi Tran}
\author[3,4]{Trung-Phuc Vo}

\affil[1]{College of Natural Sciences, Can Tho University, Can Tho City 900000, Vietnam}
\affil[2]{School of Physics and Center of Excellence on Advanced Functional Materials, Suranaree University of Technology, Nakhon Ratchasima 30000, Thailand}
\affil[3]{New Technologies-Research Centre, University of West Bohemia in Pilsen, 30100 Pilsen, Czech Republic.}
\affil[4]{Institute of Physics, Czech Academy of Sciences, Cukrovarnická 10, 162 00 Prague 6, Czech Republic.}

\affil[*]{Corresponding authors:\protect\\ 
Worawat Meevasana (worawat@g.sut.ac.th); Yen-Mi Tran (yenmi@ctu.edu.vn)}
\affil[$\dagger$]{These authors contributed equally to this work.}
\date{\today}

\begin{document}
\maketitle
\begin{abstract}
Spin–orbit coupling (SOC) plays an important role in determining the structural and electronic properties of recently proposed two-dimensional planar pentagonal materials. In this work, density functional theory calculations are employed to investigate SOC effects in p-MS$_{2}$ systems (M = Si, Ge, and Pb). Our results indicate that the p-SiS$_{2}$ structure is likely unstable, except for p-GeS$_{2}$ and p-PbS$_{2}$. A detailed j-resolved (total angular momentum) orbital analysis reveals that SOC enhances electronic localization, leading to a slight structural contraction and a reconstruction of electronic states near the Fermi level, this effect becoming stronger for heavier M atoms. While p-GeS$_{2}$ remains metallic, SOC drives a metal–semiconductor transition in p-PbS$_{2}$ and opening a quasi-direct band gap of about 0.475 eV. In addition, the conduction band minimum state of p-PbS$_{2}$ exhibits pronounced anisotropy along the S–S bonds. These findings provide insight into SOC-driven structural and electronic reconstruction in planar pentagonal chalcogenides  p-MS$_{2}$ and suggest that p-PbS$_{2}$ may be a promising candidate for gas-sensing applications.
\end{abstract}

\clearpage

\section{Introduction}
Dimensional reduction from three-dimensional crystals to two-dimensional monolayers and further to one-dimensional nanoribbons provides an effective route to engineer electronic structures through quantum confinement, reduced coordination, symmetry breaking, and modified screening. In layered transition-metal dichalcogenides, this principle is exemplified by MoS\textsubscript{2}, where thinning toward the monolayer limit reconstructs the band structure, including an indirect-to-direct band-gap crossover and spin–valley coupling relevant to optoelectronic and valleytronic applications \cite{eknapakul2014electronic}. The importance of low-dimensional electronic states is further illustrated by Meevasana and co-workers, who showed that surface-confined electron liquids in SrTiO\textsubscript{3} can be created and controlled by ultraviolet irradiation \cite{meevasana2011creation}, and that ARPES can resolve multiple interaction-driven energy scales in quasi-two-dimensional cuprates \cite{meevasana2007hierarchy}. Beyond extended 2D crystals, molecular-scale low-dimensional systems can also exhibit functionality absent in bulk materials; for example, Yang, Meevasana, and co-workers demonstrated monochromatic electron photoemission from diamondoid self-assembled monolayers, where nanoscale diamond-like molecular units produced negative-electron-affinity behavior and efficient low-energy electron emission \cite{yang2007monochromatic}. When dimensionality is further reduced to one-dimensional nanoribbons, finite-size confinement and edge chemistry can strongly modify band gaps, orbital localization, and transport channels, as shown in p-SiC\textsubscript{2} and penta-graphene-based nanoribbons \cite{mi2025sensing,mi2022comparison,mi2021diverse,mi2020adsorption,phuc2022effect,tien2019electronic,tien2020influence}. These examples demonstrate that lowering dimensionality is not merely a structural transformation, but a powerful strategy for creating electronic states and functionalities that motivate the continued exploration of new two-dimensional (2D) materials.

The field of 2D materials has expanded rapidly after the discovery of graphene in 2004 \cite{bykov2021realization, ren20252d}. Many families of 2D materials have been proposed and manufactured \cite{jindata2021spectroscopic}. A notable example is the transition metal chalcogenide family MX$_{2}$ \cite{patra2025advanced}, where M and X represent a transition metal and a chalcogen element, respectively. In these materials, chalcogen atoms form strong bonds within the atomic plane, while adjacent layers are connected by weak van der Waals forces. Therefore, MX$_{2}$ materials generally exhibit a layered structure and can be easily exfoliated \cite{patra2025advanced, liu2021mechanical, yang2022high}. Moreover, their electronic properties change from an indirect to a direct band gap when the number of layers varies \cite{hirai2025wannier}. They also exhibit high stability \cite{du2025recent, zhou2019epitaxial, hirai2025wannier} and structural diversity \cite{patra2025advanced}. However, none of them are a planar pentagonal monolayer, which is increasing in attention due to their unique properties. A representative example is penta-NiN$_{2}$ \cite{bykov2021realization}, which has been experimentally confirmed to possess a planar pentagonal lattice. Theoretical studies by Shao et al.\cite{shao2024two} further demonstrate that planar pentagonal materials can exhibit a direct band gap, high carrier mobility, anisotropic mechanical properties, and strong ultraviolet absorption. More recently, Zhang and co-workers \cite{zhang2018node} reported a class of materials similar to transition-metal chalcogenides but composed of completely planar pentagonal rings, referred to as pentagonal group-IVA chalcogenides (p-MX$_{2}$). In this structure, X (S, Se, Te) represents chalcogen elements, while M (C, Si, Ge, Sn, Pb) denotes group-IVA elements. Their work mainly focused on the influence of spin–orbit coupling (SOC) \cite{hobbs2000fully} and the atomic number (Z) on the topological-insulator properties \cite{ghiasi2025quantum} of this class of materials. Similarly, recent studies on pentagonal 2D materials \cite{shao2024two, kurpas2019spin, ono2020two, chen2021spin, zollner2025first} have primarily investigated SOC-induced changes in electronic properties. To the best of our knowledge, the influence of SOC on the structural properties and stability of p-MX$_{2}$ systems, together with a detailed analysis of electronic states based on j-resolved (total angular momentum) orbital decomposition, have not yet been reported.

In this work, we address these issues by systematically investigating the SOC-driven structural and electronic reconstruction in p-MS$_{2}$ (M = Si, Ge, and Pb). The remainder of this paper is organized as follows. The Computational Methods section describes the computational models and parameters used in our calculations. In the Results and Discussion section, we first reconstruct the p-SiS$_{2}$, p-GeS$_{2}$, and p-PbS$_{2}$ structures and analyze the influence of SOC on their geometries. The distinct structural behavior of p-SiS$_{2}$ compared to other systems is further examined through the electron density distribution and thermal stability, which indicate that this structure is unlikely to exist. The stability of p-GeS$_{2}$ and p-PbS$_{2}$ is further confirmed by comparing their cohesive energies with those of several experimental 2D materials and related systems with the same symmetry group. We then analyze the influence of SOC on the electronic properties of p-GeS$_{2}$ and p-PbS$_{2}$ using energy band structures (BS), density of states (DOS), projected density of states (PDOS), and j-resolved (total angular momentum number) orbital decomposition. Our results reveal that SOC, primarily originating from M atoms, not only induces energy-level splitting but also reconstructs the bonding characteristics and enhances the localization of electronic states near the Fermi level. In particular, SOC drives a metal–semiconductor transition in p-PbS$_{2}$, resulting in a quasi-direct band gap. It also produces distinct spatial distributions of the valence band maximum (VBM) and conduction band minimum (CBM) states along the S–S bonds in different crystallographic directions. Finally, the main findings of this work are summarized in the Conclusions section.
\section{Computational method}
Calculations are performed based on density functional theory (DFT) \cite{giustino2014materials}, as implemented in the Quantum ESPRESSO package \cite{giannozzi2009quantum, giannozzi2017advanced, giannozzi2020quantum}. The SSSP Efficiency PBEsol v1.3 pseudopotentials from the SSSP library \cite{prandini2018precision} are employed for  Si \cite{dal2014pseudopotentials}, Pb \cite{kucukbenli2014projector}, S and Ge \cite{garrity2014pseudopotentials}. To obtain optimal configurations, we set a pressure of $0.5$ kbar, a maximum force of $5 \times  10^{-4}$ Ry.Bohr$^{-1}$ on each atom, and a total energy convergence threshold of $3.0 \times 10^{-5}$ Ry. The Brillouin zone is sampled by a $6 \times  6 \times  1$ grid for geometric optimization and a $9 \times  9 \times  1$ grid for DOS calculations. The cutoff energy and the electronic energy convergence threshold are $40$ Ry and $10^{-9}$ Ry, respectively. A vacuum space of $15$ \AA {} is applied perpendicularly to the surface. VESTA software \cite{momma2008vesta} is utilized to visualize the models and their electron density distributions. The atomic positions and lattice parameters are optimized without imposing any symmetry constraints. The SOC effect is included in both the structural relaxation and the calculation of electronic properties in the ground state. Molecular dynamics (MD) simulations are performed using the FALCON on-the-fly active-learning calculator \cite{Felis2026FALCON} implemented in the Atomic Simulation Environment (ASE). FALCON employs Gaussian Process Regression (GPR) to construct a machine-learning interatomic potential and uses the predicted energy uncertainty ($\Delta E$) to identify configurations requiring DFT calculations. When $\Delta E$ exceeds the predefined threshold $\epsilon$, DFT calculations are triggered and the resulting energies and forces are added to the training set. The system is modeled using a 2 × 2 × 1 supercell and simulated in the NVT ensemble at 300 K with a Nosé–Hoover thermostat and a time step of 1 fs. The active-learning threshold is set to $\epsilon$ = 0.01 eV \cite{Felis2026FALCON}. The MD simulation is carried out for 10 ps (10$^4$ steps), and the resulting trajectory is used to evaluate the thermal stability of the structure. The formula used to calculate the cohesive energy has the following form: 
\begin{equation}
E_{c}=\frac{1}{6}\left(\sum E_{i} - E_{total}\right)
\end{equation}
where $E_{c}$, $E_{i}$, and $E_{total}$ are the cohesive energy, isolated energy of the \textit{i} atom and total energy of the optimized system, respectively.

\section{Results and discussion}
\subsection{The geometries and stability of \texorpdfstring{p-MS\textsubscript{2}}{p-MS2} (M = Si, Ge, Pb)}
The optimized structures of the p-MS\textsubscript{2} family from our calculations are shown in Figure 1. In the top view (Fig. 1a), each structure consists of connected pentagonal rings. The red dashed lines define the unit cell with lattice constants \textit{a = b}. In the side view (Fig. 1b), all atoms lie in the same plane. The structures belong to the P4/mbm space group (No. 127), as confirmed by the Bilbao Crystallographic Server \cite{aroyo2006bilbao,aroyo2006bilbao2,aroyo2011bilbao}. These structural features are in agreement with previous reports \cite{zhang2018node}. 

\begin{figure}[H]
    \centering
    \includegraphics[width=0.5\linewidth]{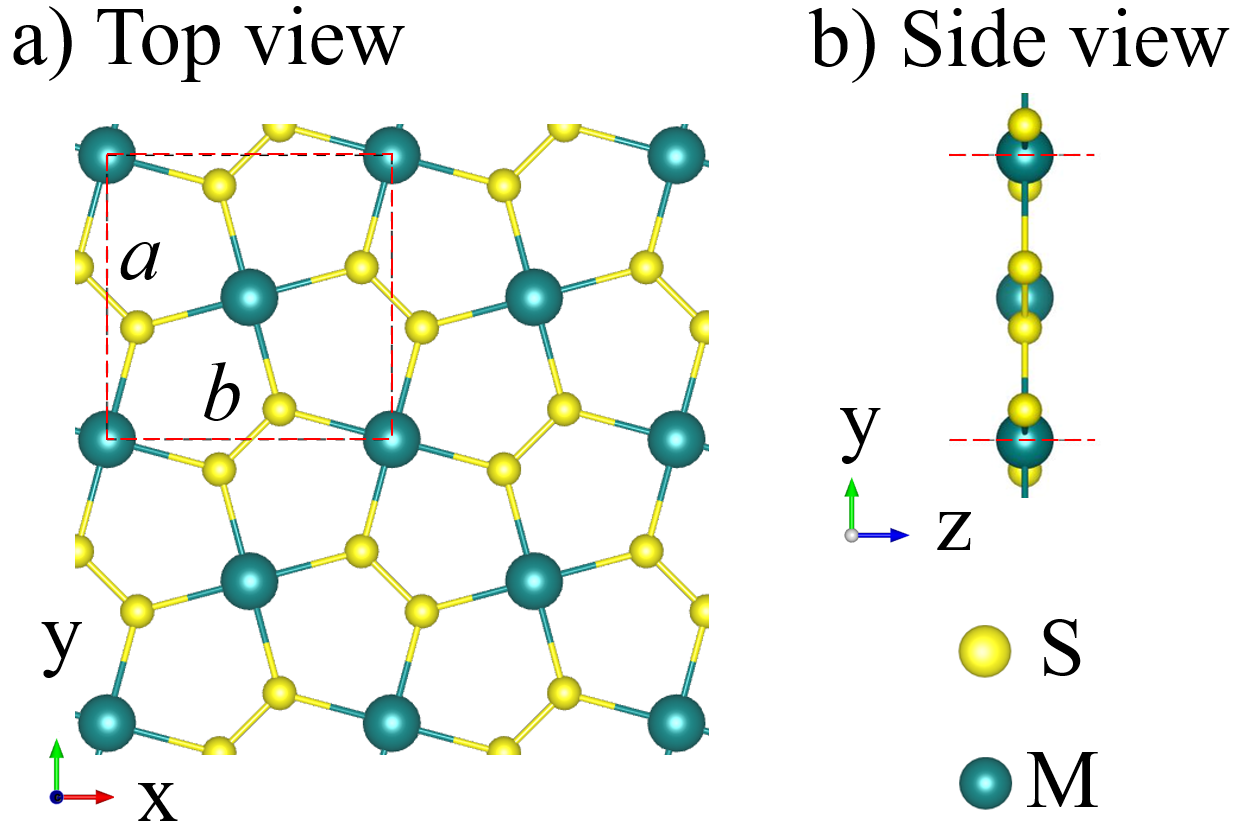}
    \caption{Optimized atomic structures of the planar pentagonal p-MS\textsubscript{2} family (M = Si, Ge and Pb). Top (a) and side (b) views are presented to illustrate the pentagonal lattice geometry and the planar configuration of the monolayers. The red dashed square in (a) defines the unit cell with lattice constants \textit{a = b}. Yellow and blue spheres denote S and M atoms, respectively.}
    \label{Figure 1}
\end{figure} 

Table I summarizes the changes in the optimized structural parameters of p-MS$_{2}$ induced by SOC. The bond lengths between nearest neighboring atoms are denoted as $d_{1}\equiv |M-S|$ and $d_{2}\equiv |S-S|$, while the ratios $\frac {d_{1SOC}}{d_{1woSOC}}$, $\frac{d_{2SOC}}{d_{2woSOC}}$ and $\frac{a_{SOC}}{a_{woSOC}}$ quantify the SOC-induced variations in $|M-S|$, $|S-S|$, and $a$. As the atomic number Z of M increases, the lattice constant increases accordingly, regardless of SOC. However, $\frac{a_{SOC}}{a_{woSOC}}$ and $\frac {d_{1SOC}}{d_{1woSOC}}$ are smaller than or equal to $100 \%$, indicating a reduction in the lattice constant and M–S bond lengths under SOC, while the S–S bond lengths remain nearly unchanged. For the p-GeS$_{2}$ and p-PbS$_{2}$ models with SOC, the M–S bond lengths are always larger than those of the S-S ones. This feature is also reported for planar penta-MX$_{2}$ structures in the P4g space group \cite{shao2024two}, and our results agree well with those of Zhang \cite{zhang2018node} (values marked with * in Table I). However, this behavior is not observed for p-SiS$_{2}$. To investigate this difference, we further analyze the electron density difference of p-SiS$_{2}$, p-GeS$_{2}$, and p-PbS$_{2}$. 

\begin{table}[htbp]
    \centering
    \resizebox{\textwidth}{!}{%
    \begin{tabular}{|c|ccc|ccc|ccc|} 
    \hline
      Sample 
      & $a_{\mathrm{woSOC}}$ 
      & $a_{\mathrm{SOC}}$ 
      & $\frac{a_{\mathrm{SOC}}}{a_{\mathrm{woSOC}}}$ 
      & $d_{1,\mathrm{woSOC}}$ 
      & $d_{1,\mathrm{SOC}}$ 
      & $\frac{d_{1,\mathrm{SOC}}}{d_{1,\mathrm{woSOC}}}$ 
      & $d_{2,\mathrm{woSOC}}$ 
      & $d_{2,\mathrm{SOC}}$ 
      & $\frac{d_{2,\mathrm{SOC}}}{d_{2,\mathrm{woSOC}}}$ \\ 
      & \AA & \AA & \% & \AA & \AA & \% & \AA & \AA & \% \\
      \hline
      p-SiS$_2$ 
      & 5.964 & $5.964/6.18^{*}$ & 100 
      & 2.200 & $2.200/2.44^{*}$ & 100 
      & 2.962 & $2.961/2.16^{*}$ & 99.9  \\
      
      p-GeS$_2$  
      & 6.335 & $6.328/6.42^{*}$ & 99.89 
      & 2.541 & $2.536/2.58^{*}$ & 99.80 
      & 2.081 & $2.085/2.08^{*}$ & 100.20  \\
      
      p-PbS$_2$  
      & 6.898 & $6.880/6.99^{*}$ & 99.74 
      & 2.815 & $2.809/2.86^{*}$ & 99.78 
      & 2.064 & $2.057/2.07^{*}$ & 99.66  \\
      \hline
    \end{tabular}%
    }
    \caption{SOC effect on the structural parameters of p-MS$_2$. The parameters $a$, $d_1=|M-S|$, and $d_2=|S-S|$ denote the lattice constant and the nearest-neighbor atom bond lengths. Subscripts SOC and woSOC include or do not include SOC, respectively. The corresponding ratios quantify the SOC-induced variations in these parameters. Values marked with $*$ are taken from Ref.~\cite{zhang2018node}.}
    \label{tab:structural_parameters}
\end{table}

Figure 2 shows the electron density difference (EDD) distribution within the unit cell of each p-MS$_{2}$. Blue regions indicate electron depletion, while red regions represent electron accumulation. Based on the electronegativity values of S ($2.58$), Pb ($2.33$), Ge ($2.01$), and Si ($1.90$), the electrons in each M–S bond tend to shift toward the S atoms. Meanwhile, the shared electron density in S–S bonds is expected to be located near the middle of each bond. However, these features are clearly observed only in p-GeS$_{2}$ and p-PbS$_{2}$. For p-SiS$_{2}$, the shared electron density along the S–S bond shifts toward one side, suggesting an unstable bonding configuration. 

To further clarify the stability of the p-MX$_{2}$ family, the thermal stability of each structure is examined using MD simulations at 300 K. Because Table 1 indicates that the SOC effect has a negligible influence on the structures of p-SiS$_{2}$ and p-GeS$_{2}$, this calculation (Fig. 3) is performed for p-SiS$_{2}$ and p-GeS$_{2}$ only without SOC, while both SOC and non-SOC cases are considered for p-PbS$_{2}$. As shown in Fig. 3a1, p-SiS$_{2}$ quickly loses its planar pentagonal structure after heating, as strong energy fluctuations occur during the first $\sim$3 ps of the simulation time and its average energy shifts noticeably. The corresponding top-view (Fig. 3a2) and side-view (Fig. 3a3)  structures further confirm its structural reconstruction at 300 K. In contrast, p-GeS$_{2}$ remains nearly stable as indicated by the small energy fluctuations ($\sim$ 0.04 eV/atom) without any noticeable energy drift (Fig. 3b1), and both its top view (Fig. 3b2) and side view (Fig. 3b3) configurations show that the structure remains intact. For p-PbS$_{2}$, the results reveal that the structure also maintains its integrity under both non-SOC (Fig. 3c1) and SOC (Fig. 3d1) conditions. However, a slight drift in its energy profile appears when SOC is included, implying that SOC may slightly reduce the stability of the system, an issue that is further discussed in later sections. Nevertheless, the structural configurations from both top (Fig. 3c2 and 3d2) and side (Fig. 3c3 and 3d3) views confirm that the lattice of p-PbS$_{2}$ remains preserved for the calculation at 300 K. Combining these results with EDD analysis as shown in Fig. 2, it could be concluded that p-SiS$_{2}$ is unlikely to be stable. 

\begin{figure}[H]
    \centering
    \includegraphics[width=1.0 \linewidth]{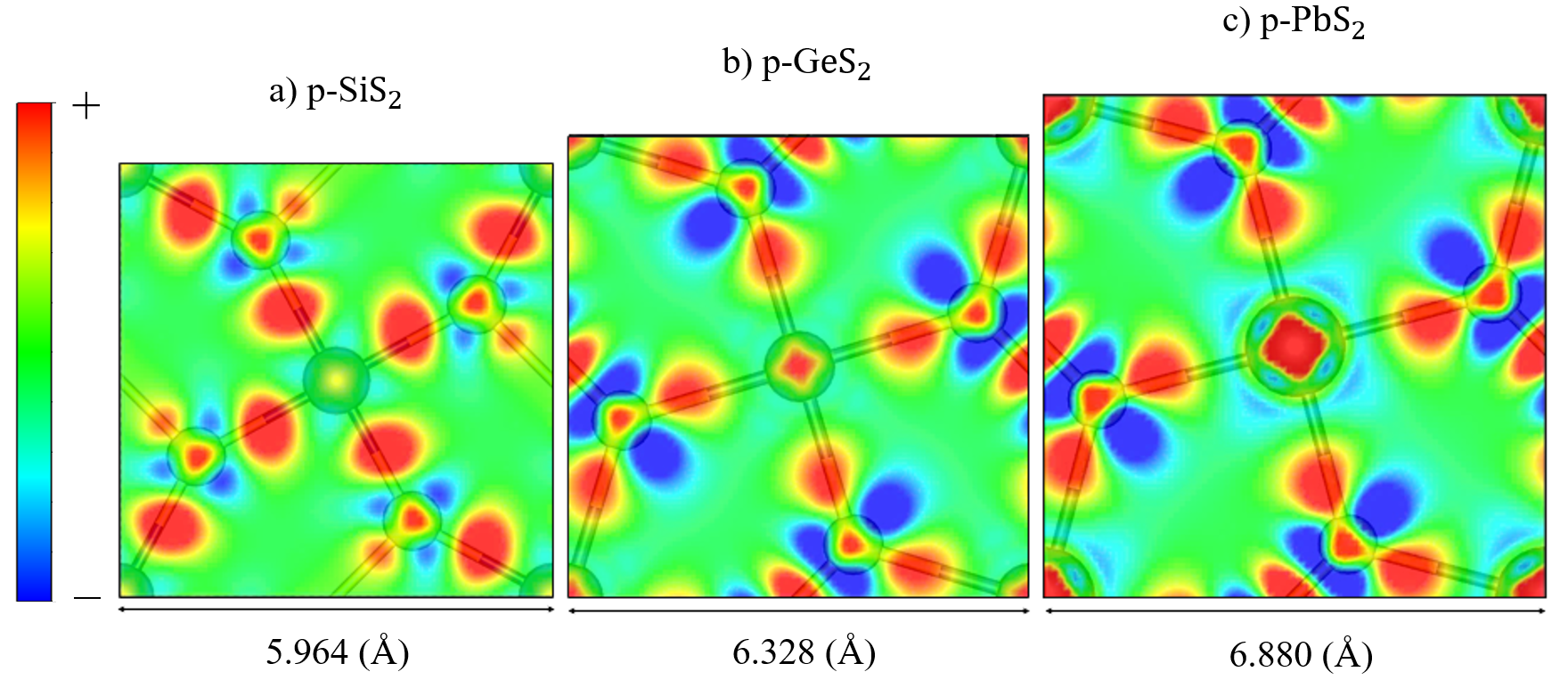}
    \caption{Electron density difference (EDD) within the unit cell of p-SiS$_{2}$ (a), p-GeS$_{2}$ (b), and p-PbS$_{2}$ (c). Blue and red regions indicate electron depletion and accumulation, respectively. The labeled structural parameters correspond to the lattice constants of each model.}
    \label{Figure 2}
\end{figure} 
\begin{figure}[H]
    \centering
    \includegraphics[width=1.0\linewidth]{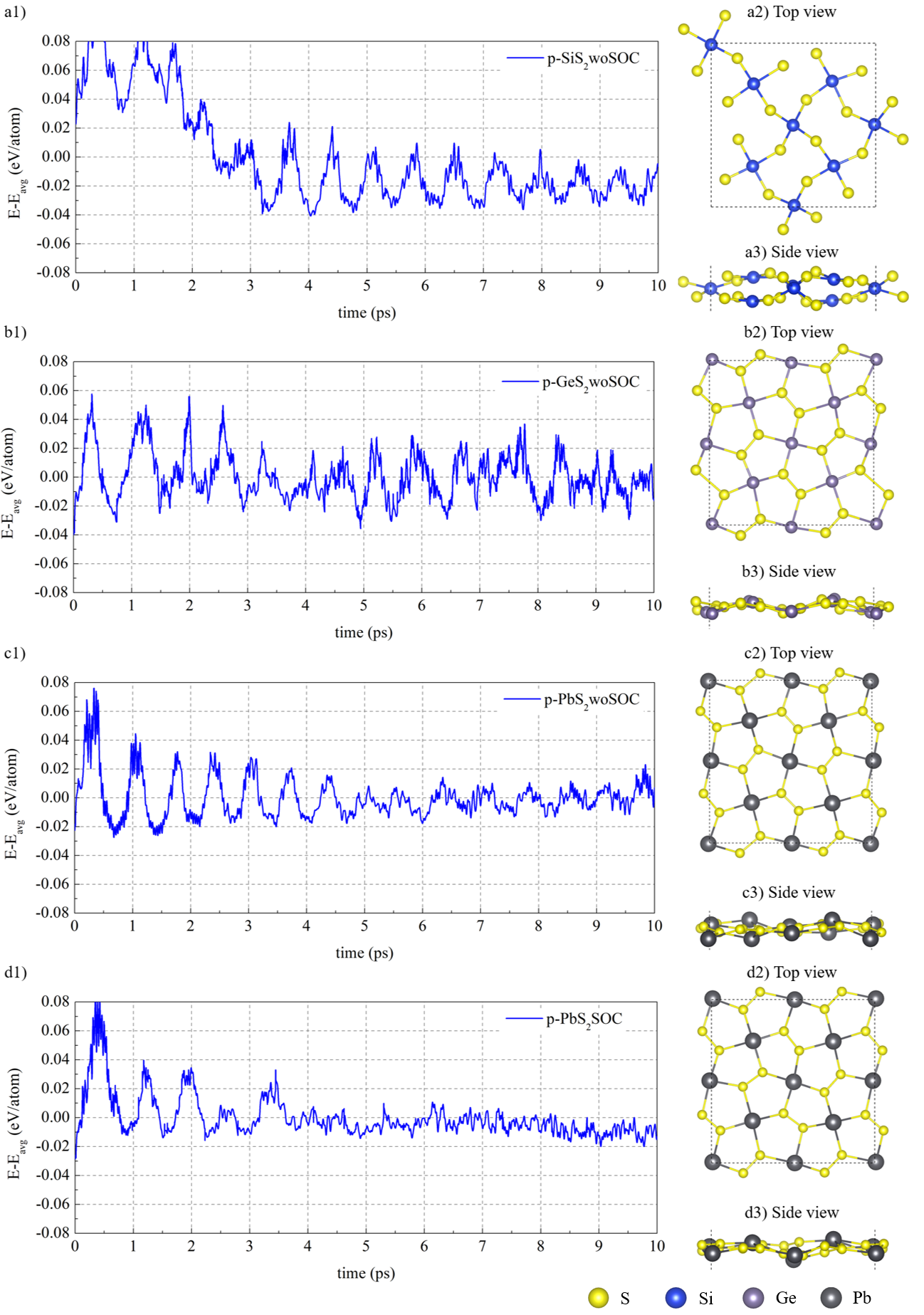}
    \caption{Total energy fluctuations as a function of simulation time during MD simulations for p-SiS$_{2}$ (a1), p-GeS$_{2}$ (b1), p-PbS$_{2}$ without SOC (c1), and p-PbS$_{2}$ with SOC (d1). The corresponding top-view and side-view structures are shown in (a2–d2) and (a3–d3), respectively. }
    \label{Figure 3}
\end{figure}

The energetic stability of p-GeS$_{2}$ and p-PbS$_{2}$ is evaluated by calculating the cohesive energy ($E_c$) using Eq. (1), defined as the difference between the total energy of the system and the sum of the energies of the isolated constituent atoms, both without and with SOC, and the corresponding ratio is used to quantify the SOC effect in their stabilities as shown in Table 2. Our results without SOC are further compared with those of other theoretical 2D materials with P4/mbm symmetry (superscript a) and experimentally realized systems (superscript b). The comparison shows that $E_c$ of several stable 2D theoretical models with P4/mbm symmetry \cite{yuan2019single} range from $4.37$ to $5.07$ eV/atom and experimental 2D materials \cite{banhart2011structural, gao2013structures, hu2015defects} exhibit values from $3.44$ to $7.50$ eV/atom, while p-GeS$_{2}$ and p-PbS$_{2}$ have cohesive energies of $4.14$ and $3.94$ eV/atom (without SOC) or 3.94 and 3.41 eV/atom (with SOC), respectively. These values fall within or close to the reported ranges, indicating that both structures can possess good energetic stability. The energy ratio for p-GeS$_{2}$ ($98.07\%$) indicates a smaller change in the stability compared to the corresponding value of p-PbS$_{2}$ ($86.55\%$) under the influence of SOC. This trend is consistent with the variation of the structural parameters shown in Table 1. By combining the analysis of structural changes and cohesive energy variations, we suggest that SOC tends to contract the lattice and decrease the stability of the p-MX$_{2}$ systems. This feature contrasts with the results reported by Shota Ono \cite{ono2020two} for square polonium monolayers, where the stability is enhanced by SOC. This discrepancy may be attributed to the presence of two atomic species in our systems, whereas their model consists of only a single element. To further elucidate these behaviors in our models, we examine the SOC-induced changes in the electronic properties of p-GeS$_{2}$ and p-PbS$_{2}$. 

\begin{table}
    \centering
    \begin{tabular}{|c|ccc|}
    \hline 
         & $E_{cwoSOC}$ & $E_{cSOC}$ & $\frac {E_{cSOC}}{E_{cwoSOC}}$ \\
        & ($eV/atom$) & ($eV/atom$) &  \\
       \hline 
       p-GeS$_{2}$  & $4.14$  & $4.06$ & $98.07 \%$ \\
       p-PbS$_{2}$  & $3.94$ & $3.41$ & $86.55 \%$ \\
       pp-Ni$_{2}$N$_{4}$  & $4.98^a$ &  &  \\
       pp-Pd$_{2}$N$_{4}$  & $4.37^a$ &  &  \\
       pp-Pt$_{2}$N$_{4}$  & $5.07^a$ &  &  \\
       Graphene  & $7.5^b$ &  &  \\
       Silicene  & $3.94^b$ &  &  \\
       Phosphorene  & $3.44^b$ &  &  \\
    \hline 
    \end{tabular}
    \caption{Cohesive energies of p-GeS$_{2}$ and p-PbS$_{2}$ calculated with and without SOC. Values with superscript a denote theoretical models in space group P4/mbm, whereas those with superscript b correspond to experimental two-dimensional materials.}
    \label{Table 2}
\end{table}

\subsection{The electronic properties of \texorpdfstring{p-GeS\textsubscript{2}}{p-GeS2} and \texorpdfstring{p-PbS\textsubscript{2}}{p-PbS2}}

The SOC-induced changes in the band structure (BS) and density of states (DOS) of p-GeS$_{2}$ and p-PbS$_{2}$ are presented in Figure 4. As shown in Fig. 4a, although SOC modifies the BS of p-GeS$_{2}$ in the $M \to \Gamma $ and $\Gamma \to X $ regions, it is not sufficient to alter its intrinsic metallic character \cite{ono2020two}. In contrast, Fig. 4b shows that SOC strongly affects p-PbS$_{2}$, transforming it from a metallic system into a semiconductor with a quasi-direct band gap of about $0.475$ eV. This pronounced effect originates from the stronger SOC associated with the heavier Pb atom, which significantly modifies the band dispersion near the Fermi level, suggesting well-defined carrier effective masses. The increasing influence of SOC with the atomic number on the electronic properties of these systems is consistent with its effects on the geometric structure and cohesive energy, as discussed above. These results are in good agreement with previous studies \cite{zhang2018node}, with the only difference being the position of the Fermi level. Similar SOC-induced modifications in the BS of other two-dimensional materials have been widely reported \cite{shao2024two, kurpas2019spin, chen2021spin, zollner2025first}.  The SOC-induced modifications in the DOS of p-GeS$_{2}$ and p-PbS$_{2}$ are further illustrated in Figs. 4c and 4d, respectively. For p-GeS$_{2}$, the effect of SOC on DOS is negligible, indicating that its electronic states near the Fermi level remain largely unchanged. In contrast, SOC has a pronounced impact on p-PbS$_{2}$, particularly in the vicinity of the Fermi level. It not only alters the electronic character of the system but also increases the localization and density of electronic states near the Fermi level. This behavior provides a clear explanation for the SOC-driven transition from metallic to semiconducting character in p-PbS$_{2}$. Moreover, the enhanced localization of electronic states is consistent with the structural contraction observed in this model, as it reduces orbital overlap and modifies the bonding characteristics.

\begin{figure}[H]
    \centering
    \includegraphics[width=1.0\linewidth]{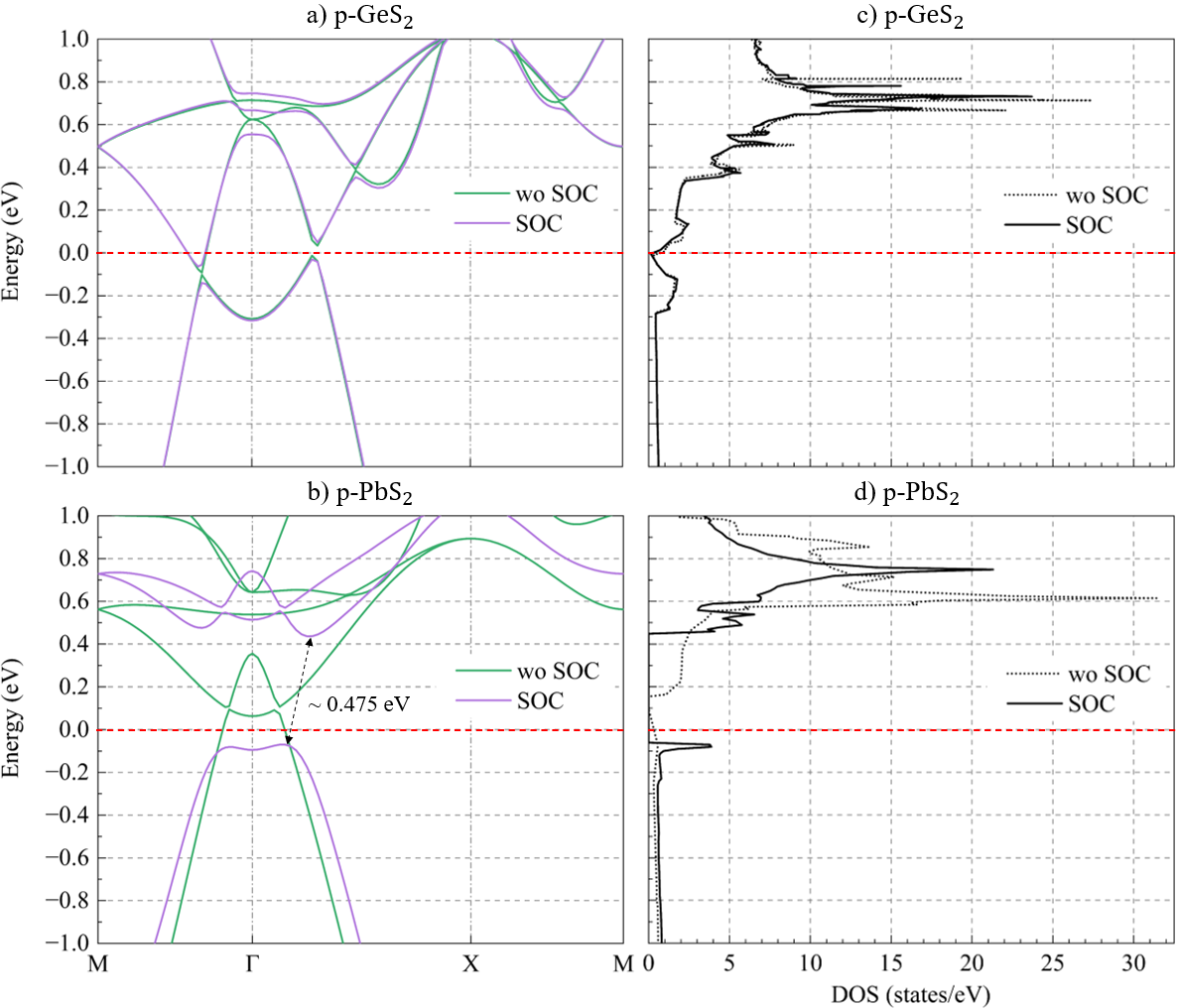}
    \caption{SOC-induced changes in the band structure (BS) and density of states (DOS) of p-GeS$_{2}$, (a) and (c), and p-PbS$_{2}$, (b) and (d). The Fermi level is set to zero energy.}
    \label{Figure 4}
\end{figure} 

To further clarify the influence of SOC on individual orbitals near the Fermi level, we analyze the projected density of states (PDOS) of p-GeS$_{2}$ and p-PbS$_{2}$, as shown in Figure 5. Figs. 5a and 5b show the negligible contribution of the d orbital of M (Pb$_{d}$, Ge$_{d}$), which can be attributed to their fully occupied nature. In contrast, the electronic states near the Fermi level are mainly dominated by the p orbitals of M (Pb$_{p}$, Ge$_{p}$) (blue lines), followed by those of S (S$_{p}$) (red lines) in both the presence and absence of SOC. With increasing Z of M, the contribution of these p orbitals becomes significantly enhanced under SOC. Notably, the appearance of the s orbital of S (S$_{s}$) (gray lines) in the VBM of p-PbS$_{2}$ (Fig. 5b) indicates that the SOC effect is sufficiently strong to modify the nature of bonding in this system. These distinct electronic characteristics suggest that p-PbS$_{2}$ is a promising candidate for gas sensing applications, with active sites likely located near the Pb atoms. 

In addition, SOC lifts the degeneracy of electronic states with orbital angular momentum quantum number $l \geq 1$. This effect is further analyzed in Figs. 5c and 5d in terms of the total angular momentum states $j = 0.5$ and $j = 1.5$ of the p orbitals. In this picture, the splitting is negligible in p-GeS$_{2}$, while a strong reconstruction of the PDOS is observed in p-PbS$_{2}$. In particular, Pb$_p$ with $j = 0.5$ shows a significant contribution at the VBM, reaching nearly $3.0$ states/eV, while the remaining states exhibit contributions comparable to those of S$_s$, as shown in Fig. 5b.  In contrast, at the CBM, the contributions from Pb$_p$ and S$_s$ ($j = 1.5$) are comparable and reach the highest values (around $2.0$ states/eV), which differs from the case without SOC. In other words, strong orbital mixing emerges, involving S$_p$ ($j = 0.5$ and $1.5$) and Pb$_p$ ($j = 1.5$) at the VBM, and S$_p$ ($j = 1.5$) with Pb$_p$ ($j = 0.5$, $1.5$) at the CBM. This pronounced hybridization further confirms that SOC modifies the nature of bonding in the system \cite{zollner2025first, gulans2022influence}, leading to a reduction in bonding strength and a decrease in structural stability. 
\begin{figure}[H]
    \centering
    \includegraphics[width=1.0\linewidth]{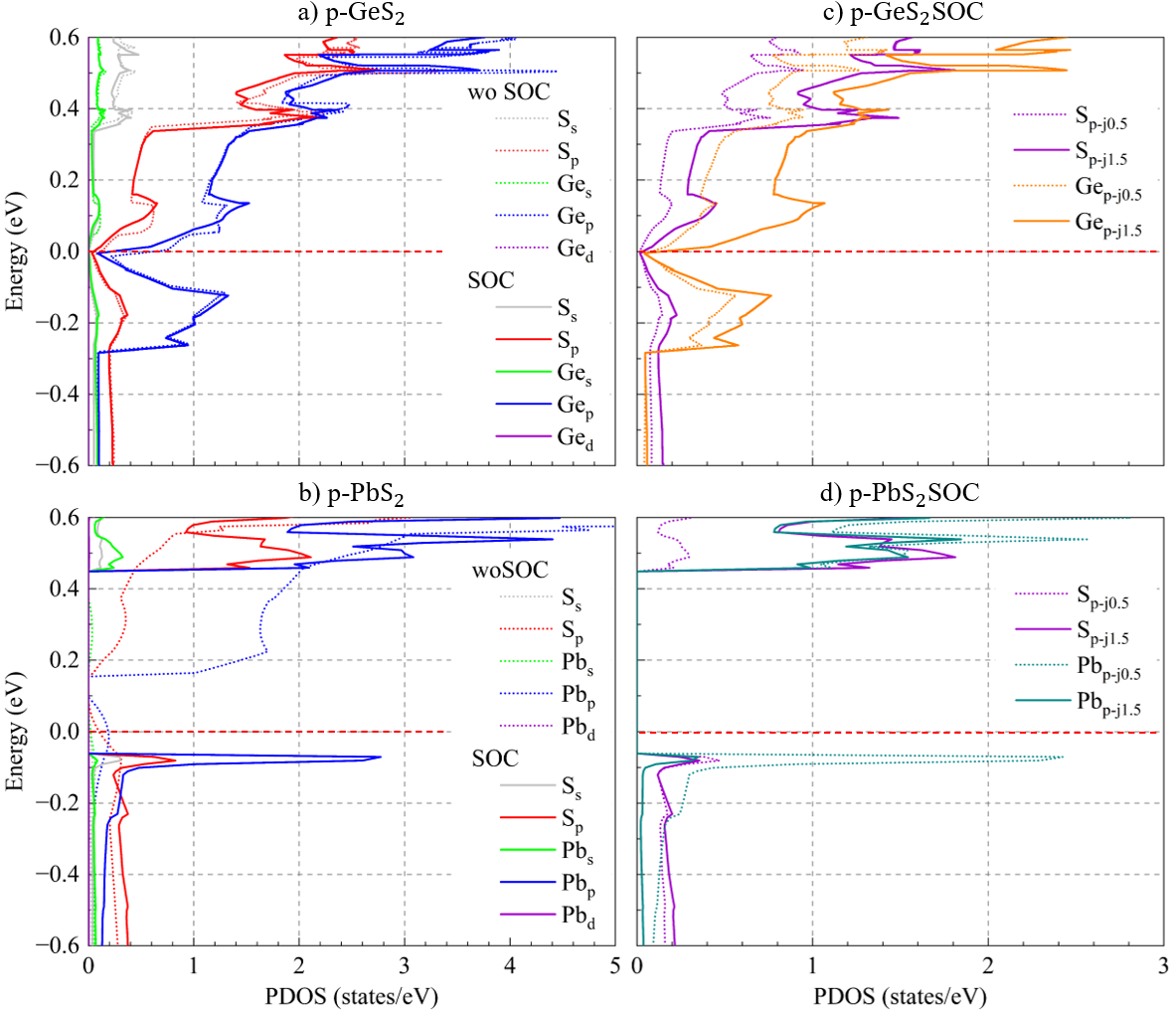}
    \caption{Projected density of states (PDOS) of p-GeS$_{2}$ and p-PbS$_{2}$ under SOC. p-GeS$_{2}$ (a) and p-PbS$_{2}$ (b) show the PDOS without decomposition into total angular momentum states, while p-GeS$_{2}$ (c) and p-PbS$_{2}$ (d) point out the $j$-resolved PDOS ($j = 0.5$ and $1.5$). The Fermi level is set to zero energy. }
    \label{Figure 5}
\end{figure} 
\begin{figure}[H]
    \centering
    \includegraphics[width=1.0\linewidth]{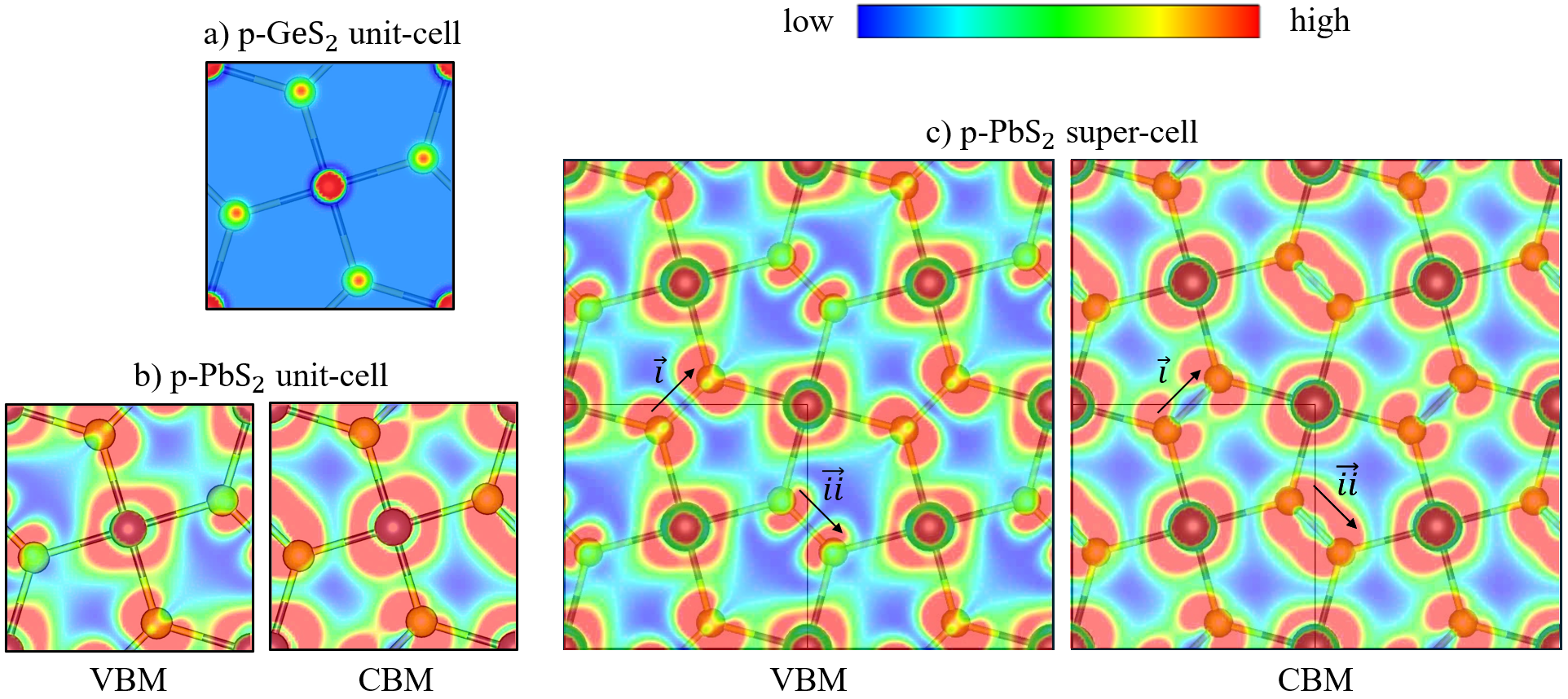}
    \caption{Spatial distribution of selected electronic states in p-GeS$_{2}$ and p-PbS$_{2}$ with SOC effect. The state at the Fermi level in the p-GeS$_{2}$ unit cell (a). The states at the VBM and CBM in the p-PbS$_{2}$ unit cell (b) and supercell (c), respectively. S-S bonding anisotropic characters are highlighted in $\vec{i}$ and $\vec{ii}$ directions.}
    \label{Figure 6}
\end{figure}

To visualize the electronic characteristics of p-GeS$_{2}$ and p-PbS$_{2}$ under SOC, the spatial distribution of selected electronic states is illustrated in Figure 6 using a cross-sectional plane that passes through all atoms. For p-GeS$_{2}$, Fig. 6a shows that the states at the Fermi level are highly localized (red regions) around the Ge atoms, while exhibiting much weaker contributions (green regions) around the S atoms within the same unit cell. This spatial distribution is in excellent agreement with the PDOS analysis presented in Fig. 5a. Fig. 6b illustrates the spatial distribution of the electronic states in the VBM and CBM within a unit cell of p-PbS$_{2}$. The states governing both band edges are predominantly localized around the atomic sites. Notably, the contribution from S atoms at the VBM is significantly weaker than that at the CBM, in excellent agreement with the PDOS analysis in Fig. 5d. Furthermore, the supercell representation of p-PbS$_{2}$ (Fig. 6c) reveals a clear anisotropy in the spatial distribution of electronic states along the S–S bonds oriented in the $\vec{i}$ and $\vec{ii}$ directions. Specifically, while the VBM states are distributed primarily along the S–S bonds with a dominant contribution along the $\vec{i}$ direction, the CBM states exhibit a significantly enhanced contribution along the $\vec{ii}$ direction and a pronounced $\pi$-bonding character emerges along this one. These results indicate that the S–S bonds along the two distinct directions within p-PbS$_{2}$ are not electronically equivalent, particularly in the CBM. This intrinsic anisotropy distinguishes p-PbS$_{2}$ from conventional pentagonal structures \cite{zhang2015penta, liu2013novel} and suggests its potential for enhanced performance in sensing applications.

\section{Conclusion}
Based on density functional theory (DFT) calculations performed using the Quantum ESPRESSO package, the structural and electronic properties of planar pentagonal p-MS$_{2}$ systems (M = Si, Ge and Pb) have been systematically investigated with the inclusion of the spin–orbit coupling (SOC) effect. Our results show that both p-GeS$_{2}$ and p-PbS$_{2}$ are energetically stable, whereas the p-SiS$_{2}$ structure is likely unstable. The inclusion of SOC leads to a slight structural contraction and marginally reduce their energetic stability. A detailed analysis based on j-resolved (total angular momentum) orbital decomposition reveals that SOC modifies the bonding characteristics by enhancing the localization of electronic states near the Fermi level. The strength of the SOC-induced effect increases with the atomic number of the M atom. While p-GeS$_{2}$ remains metallic, SOC induces a metal–semiconductor transition in p-PbS$_{2}$, opening a quasi-direct band gap of about 0.475 eV. In addition, the conduction band minimum state of p-PbS$_{2}$ exhibits a pronounced anisotropic electronic distribution along the S–S bonds. In general, these findings provide further insight into SOC-driven structural and electronic reconstruction in two-dimensional planar pentagonal chalcogenide p-MS$_{2}$ systems, particularly p-PbS$_{2}$ for sensing-related applications when SOC effects are taken into account.

\section*{Author contributions}
Phuc-Dang Truong: Writing – original draft, Investigation, Methodology, Formal analysis, Conceptualization.
Cao-Huu-Tai Nguyen: Investigation, Formal analysis.
Nguyen-Bao-Tran Ngo: Investigation, Formal analysis. 
Khanh-Van Huynh: Investigation, Formal analysis.
Yen-Mi Tran: Writing – original draft, Investigation, Supervision, Resources, Methodology, Formal analysis, Conceptualization. 
J\'{a}n Min\'{a}r: Resources, Methodology, Funding acquisition, Formal analysis, Conceptualization.
Worawat Meevasana: Resources, Project administration, Methodology, Investigation, Funding acquisition, Formal analysis, Conceptualization. 
Trung-Phuc Vo: Visualization, Validation, Supervision, Resources, Project administration, Methodology, Investigation, Formal analysis, Conceptualization. All authors contributed to the writing, review, and editing of the manuscript.

\section*{Acknowledgements}
This work was supported by the Program Management Unit for Human Resources and Institutional Development, Research and Innovation (Thailand)[Grant Numbers B39G680007 and B13F670064], Suranaree University of Technology, Thailand Science Research and Innovation (TSRI) and National Science Research and Innovation Fund (NSRF). T.-P.V. acknowledges the project Quantum materials for applications in sustainable technologies (QM4ST), funded as project \texttt{No. ~\nolinkurl{CZ.02.01.01/00/22\_008/0004572}} by Programme Johannes Amos Commenius, call Excellent Research and the Czech Science Foundation Grant No. GA \v{C}R 23-04746S.

\section*{Ethics declarations}
\subsection*{Competing interests}
The authors declare no competing interests.

\bibliography{TLTKp-MS2}
\end{document}